\documentclass[a4paper]{jpconf}
\usepackage{graphicx}

\newcommand{\be}{\begin{equation}}
\newcommand{\ee}{\end{equation}}
\newcommand{\ba}{\begin{array}}
\newcommand{\ea}{\end{array}}
\newcommand{\bea}{\begin{eqnarray}} 
\newcommand{\eea}{\end{eqnarray}} 
\newcommand{\bd}{\begin{displaymath}}
\newcommand{\ed}{\end{displaymath}}
\newcommand{\eps}{\varepsilon}

\newcommand{\trm}[1]{\textrm{#1}}

\newcommand{\figref}[1]{Fig. \ref{#1}}

\newcommand{\eqnref}[1]{Eq. (\ref{#1})}

\newcommand{\vphi}{\varphi}

\newcommand{\sech}{\trm{sech}}

\begin{document}
\title{Beyond Volkov: Solving the Second-Order Klein-Gordon Equation}

\author{B. King}
\address{Centre for Mathematical Sciences, Plymouth University, PL4~8AA, UK}
\ead{b.king@plymouth.ac.uk}

\begin{abstract}
Whether monochromatic, pulsed, or even constant and crossed, the field used to describe the interaction
of charged fermions with an intense laser beam is mainly assumed to be of plane-wave form. We consider a simple extension to plane-wave fields and consider a scalar particle in a non-lightlike, univariate and transverse propagating electromagnetic wave. The existence of some known exact solutions in this case allows
us to analyse various proposed approximations in the literature as well as the plane wave model. The results also describe some of the quantum dynamics of a scalar particle in a standing wave background.
\end{abstract}

\section{Introduction}
We wish to calculate the behaviour of an electron in a realistic external electromagnetic (EM) field. There are few classes of external field, for which the Dirac equation has been solved analytically. One example is the solution to the Dirac equation in a plane wave background, the so-called ``Volkov solution'' \cite{volkov35}. This solution is central to the \emph{plane wave model} of laser-based strong-field quantum electrodynamics (SFQED) and has dominated calculations for the last few decades (more detail on SFQED can be found in reviews \cite{ritus85,marklund_review06,gies09,dipiazza12,narozhny15,king15a}). However, to acquire the high field intensities in experiment, a laser beam has structure in both space, via focussing, and time, via pulse compression, so is clearly not a plane wave. To assess whether the plane wave model is a good approximation for SFQED processes in highly intense laser pulses, one should be able to find the plane wave limit from more complicated backgrounds and investigate its domain of applicability.
\newline

% The pulse has a finite bandwidth, and so when the field is squared, there 
% Since the pulse has a finite frequency bandwidth, the assumption of it being a plane wave may be questionable. For example, suppose 

The standard argument of the plane wave model is the following \cite{ritus85}. QED is a covariant theory, so the probability of any process $P_{\tiny{\trm{QED}}}$ can be written entirely in terms of relativistic invariants. The natural field scale is given by the ``Schwinger'' field $E_{\tiny{\trm{cr}}} = m^{2}c^{3}/e\hbar$, where $m$ and $e$ are the mass and charge of a positron (from here on, we set $c$, the speed of light in vacuo and $\hbar$, Planck's constant to the values $c=\hbar=1$). Then the relevant field invariants are the usual EM invariants, scaled by the Schwinger field $\mathcal{F} = -e^{2}F^{\mu\nu}F_{\mu\nu}/4 m^{4}$ and $\mathcal{G} = -e^{2}F^{\mu\nu}F^{\ast}_{\mu\nu}/4 m^{4}$, where $F_{\mu\nu}$ and $F^{\ast}_{\mu\nu}$ are the Faraday tensor 
and its dual \cite{jackson75}. Two further invariants can be defined. The \emph{intensity parameter} $\xi = e \sqrt{\langle p\cdot T(\vphi)\cdot p \rangle_{\vphi}} / m\,(k\cdot p)$ \cite{ilderton09} where $T^{\mu\nu}=(F^{2})^{\mu\nu} - \eta^{\mu\nu}\, \tr F^{2} / 4$ is the energy-momentum tensor \cite{jackson75}, $\langle\cdot\rangle_{\vphi}$ indicates a cycle-average over $\vphi$, $p$ is the electron momentum and $k$ is the external-field wavevector, quantifies the work done by the external field over a Compton wavelength in units of a the field's photon energy. The \emph{energy parameter} $\eta = k\cdot p/m^{2}$ quantifies the seed particle energy. Most discussions of the plane wave model choose the quantum parameter $\chi = \eta \xi$ instead of $\eta$, so we choose that here as well. Then probabilities in QED can be written $P_{\tiny{\trm{QED}}} = P_{\tiny{\trm{QED}}}(\xi,\eta,\mathcal{F},\mathcal{G})$. Typical laser intensities (to the best of our knowledge the highest recorded is $2\times 10^{22}\,\trm{Wcm}^{-2}$ at the HERCULES laser \cite{yanovsky08}), are much less than the equivalent Schwinger intensity for a linearly-polarised pulse: $I_{\tiny{\trm{cr}}} = 4.6\times 10^{29}\,\trm{Wcm}^{-2}$, and so one can assume $\mathcal{F}$, $\mathcal{G}$ are the smallest parameters. This allows one to Taylor-expand probabilities in these parameters and assuming they enter only perturbatively, or that their non-perturbative contribution is vanishingly small, $P_{\tiny{\trm{QED}}}(\xi,\xi\,\eta,\mathcal{F},\mathcal{G})\approx P_{\tiny{\trm{QED}}}(\xi,\xi\,\eta,0,0)$. This implies that when $\mathcal{F},\mathcal{G}\ll \xi, \xi\,\eta$ and $\mathcal{F},\mathcal{G} \ll 1$, it is a good approximation to assume an arbitrary background is \emph{crossed} (electric and magnetic fields perpendicular and equal in magnitude). A focussed laser pulse is then taken to be a perturbation around the plane wave background which is an example of a propagating crossed field.
\newline

The plane wave model is central to numerical codes that wish to include QED effects in laser-plasma interactions \cite{ruhl10,elkina11,king13a,ridgers14,mironov14,gonoskov15,gelfer15,harvey15b,king16}. This is because they rely upon the locally constant field approximation. This has the premise that when $\xi \gg 1$, formation regions of QED processes become much smaller than the pulse wavelength and so the field can be approximated as locally constant \cite{harvey15}. Therefore a better understanding of when the accuracy of the plane wave model is questionable will also have an impact on experimental design and analysis, which invariably invokes numerical simulation.
\newline

The Proceeding is organised as follows: we begin in section $2$ with a recap of non-lightlike fields, followed by an analysis of solutions to the Klein Gordon equation for $k^{2}<0$ in section $3$, detailing scalar charged particle dynamics for the case of over-the-barrier, under-the-barrier and periodic background scattering. In sections $4$ and $5$ we analyse the solution and various approximations by calculating the scalar particle quasimomentum and the field-theory current. In section $6$ we conclude the presentation of results.

\section{Non-lightlike fields}
The Faraday tensor of a plane wave can be written $F^{\mu\nu} = F^{\mu\nu}(\vphi)$ where $\vphi = k\cdot x$ and $k^{2}=0$. We choose to relax one of these conditions and study non-lightlike fields, for which $F = F(\vphi)$ but $k^{2} \neq 0$. If $k^{2}>0$, then one can perform a Lorentz transformation to a frame in which $k = \omega(1,0,0,0)$ and the wave is entirely \emph{timelike}. This would correspond to a homogeneous but time-dependent electric field. This case has been studied in various works \cite{becker77,mendonca11,raicher13,varro13,varro14,varro14b,raicher15,raicher16}. If $k^{2}<0$ then one can perform a Lorentz transformation to a frame in which $k = \omega(0,0,0,1)$ and the wave is entirely \emph{spacelike}. This would correspond to an inhomogeneous but constant magnetic field. This has also been studied in various works \cite{becker77,cronstroem77}. Some motivation for studying SFQED in non-lightlike backgrounds also comes from experiments using energetic particle beams with oriented crystals \cite{uggerhoj5}, suggested experiments using laboratory plasmas \cite{dipiazza07} and observations of strongly-magnetised astrophysical systems \cite{harding06}.
\newline

If one takes the magnetic case $k = \omega(0,0,0,1)$, then after performing a Lorentz transformation in the $z$-direction, the wavevector becomes $k = \omega\gamma \beta(1,0,0,1/\beta)$. As $\beta \to 1$, the wavevector tends to that of a plane wave background. But however relativistic the transformation, $k^{2} \neq 0$, as it is invariant. We highlight that non-lightlike fields are relativistically inequivalent to plane wave fields. In particular, charges in undulators and charges in lasers are not equivalent \cite{harvey12}.
\newline

Combinations of plane-wave fields can be both of magnetic and electric character. Take a standing wave made from two, counterpropagating, circularly-polarised plane waves, which has a vector potential:
\[
 A(\vphi_{1},\vphi_{2}) = C\left\{\eps_{1}\cos\vphi_{1}+\eps_{2}\sin\vphi_{1} + \eps_{1}\cos\vphi_{2}+\eps_{2}\sin\vphi_{2}\right\},
\]
where $\{k_{1}, \eps_{1}, \eps_{2}\}$ and $\{k_{2}, \eps_{1}, \eps_{2}\}$ are two \emph{dreibeins} and we pick $k_{1} = \omega(1,\vec{n})$, $k_{2} = \omega(1,-\vec{n})$, with $\vec{n}\cdot\vec{n} = 1$. Ponderomotive terms in a plane wave are related to the square of the vector potential. We see:
\[
 A^{2}(\vphi_{\Delta}) = -2C^{2}(1+ \cos \vphi_{\Delta}); \qquad \vphi_{\Delta} = k_{\Delta}\cdot x, \quad k_{\Delta} = k_{1}-k_{2},
\]
and $k_{\Delta}^{2} < 0$. Alternatively, at a magnetic node of this standing wave, for example if $\vec{n} = (0,0,1)$, in the $z=0$ plane, we see $\vphi_{1}=\vphi_{2} = \omega(1,0,0,0)\cdot x=  \bar{k}\cdot x$ where $\bar{k} = \omega(1,0,0,0)$, for which $\bar{k}^{2} >0$. Classical and quantum electron dynamics for these two cases have recently been studied in \cite{king16c}, in the following, we concentrate on a single plane wave with magnetic character $k^{2}<0$.

\section{The second-order Klein-Gordon equation}
In recent publications \cite{king16c,king16b}, the classical and scalar quantum dynamics in some example non-lightlike backgrounds have been analysed. Here we concentrate on the solution of the Klein-Gordon equation:
\[
\left[D^{2}+m^{2}\right]\Phi = 0; \qquad D = \partial + ia; \qquad a = eA.
\]
One can proceed with the usual ansatz that is employed to acquire the plane wave solution $\Phi = F(\vphi)\exp(i p\cdot x)$, to give:
\[
 k^{2}\,F'' - 2i\,k\cdot p\,F' + (2\,a\cdot p - a^{2})F = 0.
\]
If the plane-wave case is taken and $k^{2}\to0$ (not via a boost, but via a co-ordinate rotation \cite{hornbostel91,ji01,ji13,ilderton15}) set, we see that the solution can be solved immediately by exponentiation and we recover:
\[
 F = \exp \left[-i u_{\tiny\trm{pw}}(\vphi)\right]; \qquad u_{\tiny\trm{pw}}(\vphi) = \int^{\vphi} \frac{2 a(\phi)\cdot p - a^{2}(\phi)}{2k\cdot p}d\phi,
\]
where $u_{\tiny\trm{pw}}$ is the plane wave or ``Volkov'' exponent \cite{volkov35}. However, if $k^{2}\neq 0$, the Klein-Gordon (KG) equation is of second-order. We immediately identify that a \emph{perturbative} approach of neglecting the $k^{2}$ term is problematic: a condition for perturbation theory to apply is that the number of solutions to an equation should not change in the limit of of zero perturbation \cite{bender78}. Still, one of the solutions will turn out to be unphysical, so there is some hope the plane wave limit may still be a good approximation. Making the alternative ansatz $\Phi = G(\vphi)\exp(i \tilde{p}\cdot x)$ where $\tilde{p} = p - (k\cdot p/k^{2})k$, we acquire a second-order equation for $G$ in normal form:
\bea
 k^{2} G'' + 2(a\cdot \tilde{p} - a^{2})G = \left(\tilde{p}^{2}-m^{2}\right)G. \label{eqn:G}
\eea
Since we are interested in the plane-wave limit and perturbations around this, $k^{2}$ can be the smallest invariant in the problem. This motivates us to map \eqnref{eqn:G} onto the  nonlinear Schr\"odinger equation:
\bea
 -\frac{\hbar^{2}}{2}G'' + V(\vphi) G = \mathcal{E} G, \label{eqn:S}
\eea
for potential $V$ and energy $\mathcal{E}$. We distinguish three cases: i) over the barrier; ii) under the barrier scattering and iii) a periodic background.

\subsection{Over the barrier / head-on scattering}
Let us consider $k^{2}<0$ and $a = m\xi \hat{f}_{\mu}$ where $\hat{f}_{\mu}$ is of order unity. $k^{2}$ is the smallest parameter, and to set the scattering to be over the barrier, we pick $k\cdot p/\sqrt{-k^{2}}\gg p_{\perp}$, where $p_{\perp}$ is the electron's momentum perpendicular to the external field wavevector. We also assume $p_{\perp} \ll m \xi$ (the transverse momentum acquired by an electron in a plane wave field is of the order $p_{\perp} \approx m \xi$ \cite{ilderton09}). Normalising \eqnref{eqn:G} by $(m\xi)^{2}$, we identify:
\[
 V \sim \hat{f}^{2} \sim 1; \qquad -\frac{\hbar^{2}}{2}  = \frac{k^{2}}{(m\xi)^{2}} \ll 1.
\]
For this to be over-the-barrier scattering, we require the particle energy to be much larger than the potential. In other words, we require $\mathcal{E}\sim (\gamma/m\xi)^{2} \gg 1$. This ``high-energy'' set-up then ensures the smallest parameter is multiplying the highest derivative, which is the typical situation where one can employ the WKB method to acquire a semiclassical solution \cite{bender78}:
\bea
G(\vphi) \sim \left[\frac{1}{\mathcal{E}-V(\vphi)}\right]^{1/4}\exp\left[\pm\frac{i}{\hbar} \int^{\vphi} \sqrt{2(\mathcal{E}-V)}\right] \label{eqn:W}
\eea
(the $\pm$ sign will be fixed by the $\xi \to 0$ limit or the asymptotic sign of $k\cdot p$). This approach is known in SFQED \cite{blankenbecler87,baier94} and has recently been further developed to study high-energy particle-laser collisions \cite{dipiazza13b,dipiazza15,dipiazza16}.

\subsection{Under the barrier / wide-angle scattering}
\begin{figure}[t!!]
\centering
 	\includegraphics[width=0.7\columnwidth]{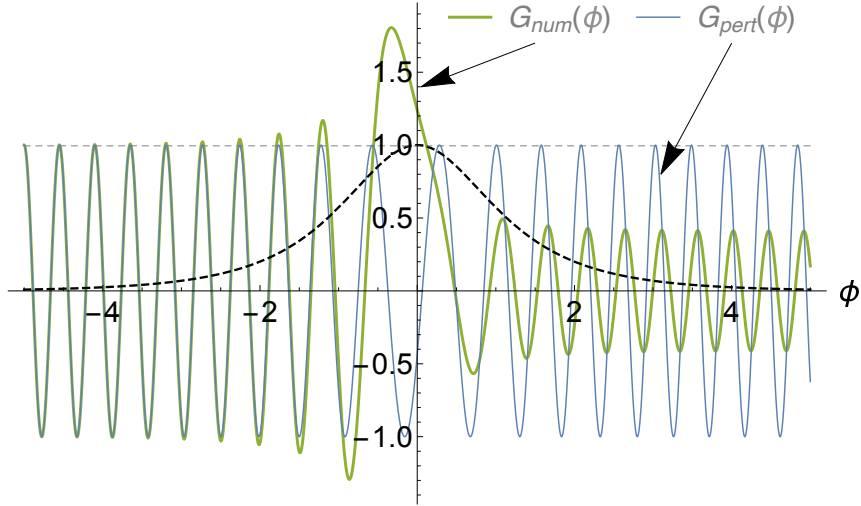}
	\caption{\label{FIG:SECH} Comparison of the numerical solution of the Schr\"odinger equation 
in the sech-type potential with the perturbative approximation of neglecting the $k^{2}$ term in the KG equation. The particle energy $\mathcal{E}=0.995$ 
(horizontal dashed line) is chosen to be just below the peak of the potential (black dashed line). The initial conditions
initial conditions $G(-5)=1$, $G'(-5)=0$ have been used.}
\end{figure}

In contrast to the high-energy, head-on scattering case, this type of dynamics is characterised by lower energies and large transverse momenta. It is known that quantum effects can lead to a transverse spreading of an electron beam that is distinct from classical or beam-shape effects \cite{harvey14}. To demonstrate barrier effects, we take the background field to be of the form: $a_{\mu} = m \xi l_{\mu} \sech(\vphi) = m \xi l_{\mu} g(\vphi)$, which produces a $\trm{sech}$-like localised potential maximum. In this case we have:
\[
 -\frac{\hbar^{2}}{2} = \frac{k^{2}}{3m^{2}\xi^{2}}; \quad V = \frac{g^{2} + 2g}{3}; \quad \mathcal{E} = \frac{\tilde{p}^{2}-m^{2}}{3m^{2}\xi^{2}},
\]
and we have chosen the electron's perpendicular momentum to be of the order of that acquired in a plane-wave background $p_{\perp} = -m\xi l_{\perp}$. We also take $\tilde{p}^{2} \approx m^{2}$. We then pick $\mathcal{E} = 0.995$ for a potential of height $V=1$, to demonstrate barrier transmission. In \figref{FIG:SECH} the reflection and transmission of the electron wavefunction becomes evident. Also plotted is the perturbative, plane-wave-like result, which is effectively blind to the barrier.

\subsection{Periodic fields}
\begin{figure}[t!!]
\centering
	\includegraphics[width=8cm]{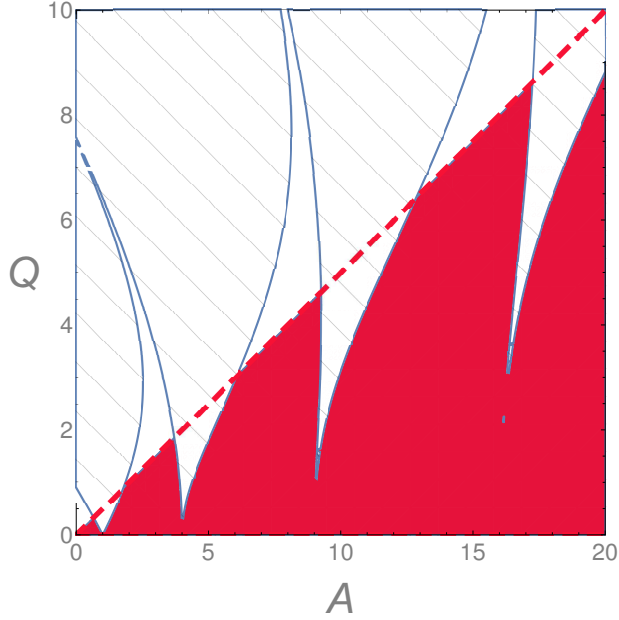}
	\caption{\label{fig:floquet} A plot of the Floquet exponent. Forbidden parameter regions or ``gaps'', in which the imaginary part of the Floquet exponent is non-zero, are indicated by linear hatching. Permissible regions or ``bands'', in which the imaginary part of the Floquet exponent is zero  ``bands'' are indicated by solid colours. The shaded area within the dotted line is in principle accessible to an electron in a circularly-polarised monochromatic plane wave for which $k^{2}<0$.}
\end{figure}

If the KG equation is written for monochromatic, circularly-polarised vector potential, it can be cast in the form:
\[
 \frac{d^{2}G}{dy^{2}} - 2Q\cos(2y)\,G = -A\, G; \qquad A = \frac{4}{k^{2}}\left(\frac{(k\cdot p)^{2}}{k^{2}} + m^{2}\xi^{2}\right); \quad Q = -\frac{4m\xi|p_{\perp}|}{k^{2}},
\]
and $y = \vphi/2$, which is a recognised form of the \emph{Mathieu Equation} \cite{bender78} (the case of linear polarisation leads to a Hill equation, recently studied in \cite{varro13,varro14,varro14b}). Solutions to the Mathieu equation can be written in the form $G = \phi(\vphi) \exp[i\nu(A,Q)\vphi]$, where $\phi$ is a periodic function and $\nu(A,Q)$ is referred to as the ``Mathieu characteristic exponent'' \cite{bender78} or ``Floquet exponent'' \cite{muellerkirsten06}. Certain regions of $A$-$Q$ parameter-space lead to an imaginary Floquet exponent and since $\vphi$ can be arbitrarily large and of positive or negative sign, this would indicate an infinitely large wavefunction normalisation constant. These regions are referred to as ``gaps'' due to the vanishing probability of an electron possessing these parameters for an arbitrarily-long time. Regions where the imaginary part of the Floquet exponent are zero are referred to as ``bands''. (A recent discussion of the connection of this band structure with resurgence can be found in \cite{dunne16} and references therein.) The position of these structures is indicated in \figref{fig:floquet}. 

Also the Mathieu equation can be mapped onto the Schr\"odinger equation in \eqnref{eqn:S} using the following assignment:
\[
 \frac{\hbar^{2}}{2} = \frac{2}{Q}, \quad V = \cos \vphi, \quad \mathcal{E} = \frac{A}{2Q}.
\]
The cosine potential describes an infinite number of degenerate local minima. The large-$Q$ limit corresponds to an electron being captured in a minimum which locally resembles a harmonic oscillator with high and steep walls. In the small-$Q$ limit, one expects tunneling between neighbouring minima to become more probable.

\section{Quasimomentum}
\begin{figure}[t!!]
\centering
 	\includegraphics[width=7cm]{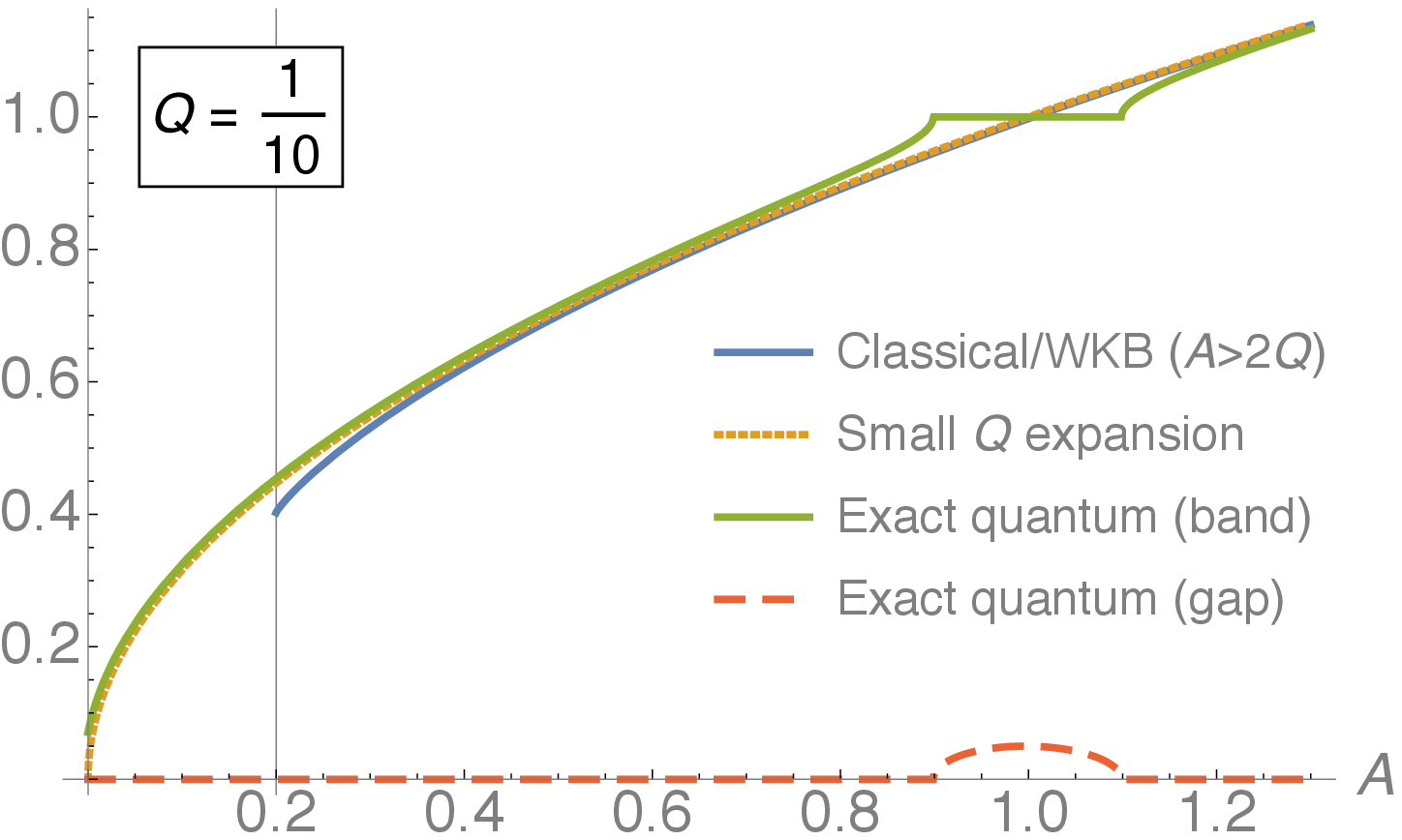} \hfill \includegraphics[width=7cm]{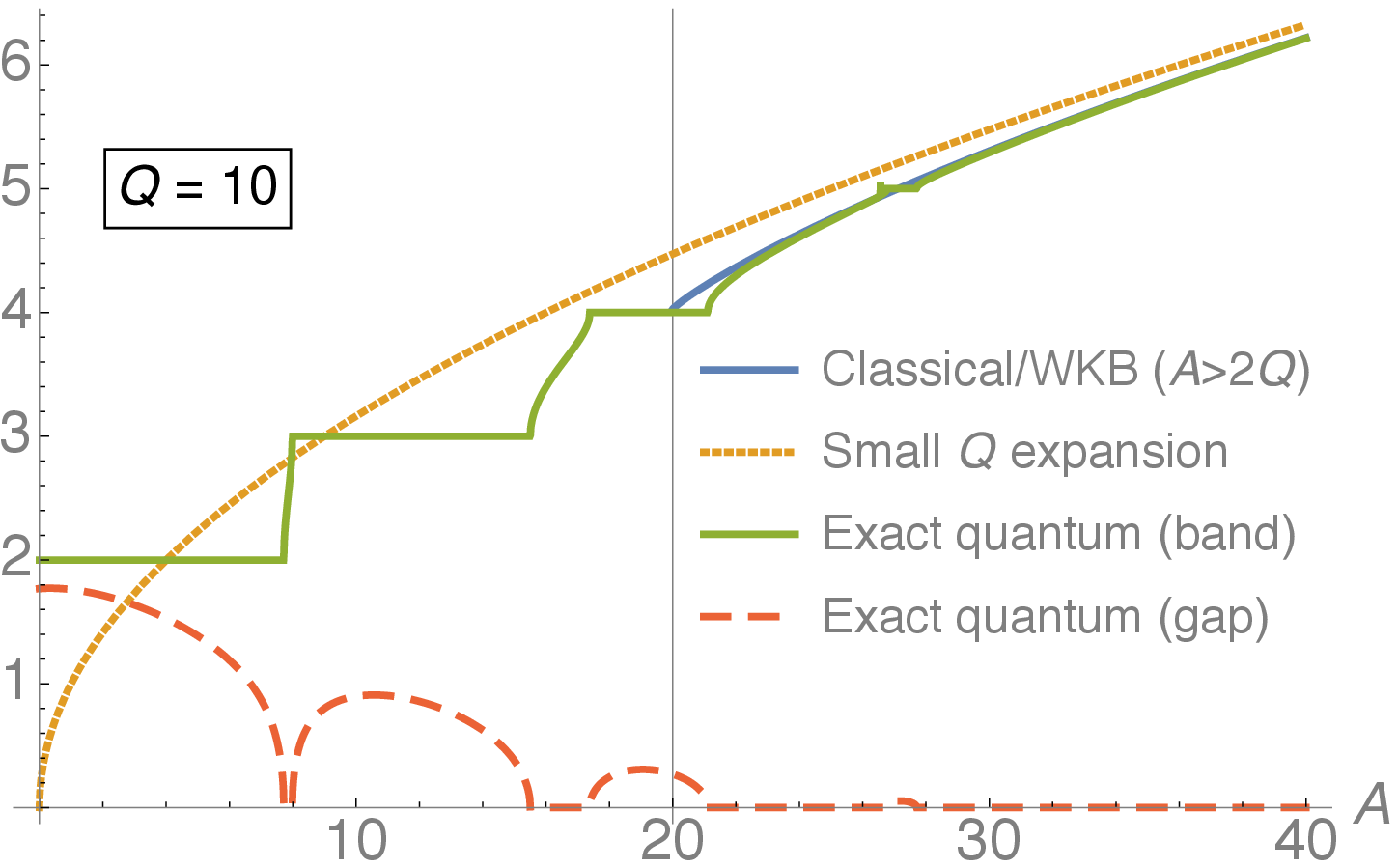}
	\caption{\label{fig:nup} Comparison of the longitudinal quasimomentum component for various approximations. Left: $Q=1/10$ small-$Q$ example. Right: $Q=10$ large-$Q$ example. The imaginary parts of the exact solution are shown with orange dashed lines, whereas the real part is shown with a solid green line. Left of the vertical line delineates the under-the-barrier, classically-forbidden region. Horizontal lines in the exact solution indicate forbidden regions. For $A < 2Q$, under the barrier, the
band and gap structure is clearly visible. However, the approximations are blind to the band/gap structure, which also occurs at values over the barrier. 
}
\end{figure}

To compare various approximations to the periodic background solution, one can study the \emph{quasimomentum} $q$ in each approach. The quasimomentum is that quantity which occurs in global energy-momentum conserving delta-functions. For the $k^{2}<0$ periodic background case, the exact quasimomentum can be written as \cite{becker77,cronstroem77} $q_{\mu} = \tilde{p}_{\mu} -\trm{sign}(k\cdot p)\,\nu(A,Q)\,k_{\mu}/2$. The exponent in the WKB case \eqnref{eqn:W} contains an incomplete elliptic integral of the second kind, $\trm{E}(\cdot|\cdot)$, leading to a quasimomentum similar in form to the exact quasimomentum, but with $\nu(A,Q)\to \sqrt{A-2Q}\,\trm{E}[\pi/2,-4Q/(A-2Q)]$ when the cycle-average is performed. If $Q$ is small, the exact quasimomentum can be written using the small-$Q$ expansion of the exponent $\nu(A,Q) \approx \sqrt{A}$ \cite{muellerkirsten06}. The accuracy of these approximations as well as the classical solution for the longitudinal component of the quasimomentum, are displayed in \figref{fig:nup}.

\section{Current conservation}
One way to understand the discrepancy between the perturbative, plane-wave-like result and the correct electron wavefunction is to calculate the field theory current:
\[
 J_{\mu}(x) = \Phi^{\dagger}\partial_{\mu}\Phi - \partial_{\mu}\!\left(\Phi^{\dagger}\right)\Phi + 
 2 a_{\mu} \Phi^{\dagger}\Phi.
\]
From current conservation, we know $\partial^{\mu}J_{\mu} = (d/d\vphi) k\cdot J = 0$. Then we find:
\[
 J_{\mu}^{\tiny\trm{pw}} = p_{\mu} -a_{\mu}(\vphi) + u_{\tiny\trm{pw}}(\vphi)\,k_{\mu}; \qquad J_{\mu}^{\tiny \trm{wkb}} = \frac{\pi_{\mu}(\vphi)}{s(\vphi)}
\]
where we have used the definition
\[
 \pi_{\mu} = p_{\mu} - a_{\mu}(\vphi) + \frac{k\cdot p}{k^{2}}\left[s(\vphi)-1\right]\,k_{\mu}\qquad s(\vphi) = \sqrt{1+\frac{2k^{2}}{k\cdot p}\,u_{\tiny\trm{pw}}}.
\]
Since $\vphi = k\cdot p$, we see that for WKB, and hence classically, the current is conserved as $k\cdot J^{\tiny\trm{wkb}}/ k\cdot p = 1$. However, for the plane wave model we have: $k\cdot J^{\tiny \trm{pw}} = 1 + 2\eps u(\vphi)$, for $\eps = k^{2}/2k\cdot p \ll 1$. Therefore, the plane wave model violates current conservation to the order $O(\eps)$, which is particularly relevant for cases of nontrivial barrier transmission/reflection, as already shown in \figref{FIG:SECH}.

\section{Conclusions}
The plane wave model has been the focus of laser-based SFQED calculations for several decades. Due to a linear relationship between the electron phase and and the proper time, the classical dynamics in a plane wave is integrable (there exist three conserved quantities in addition to the mass-shell condition). The quantum case is essentially the WKB solution, which is exact. However, when the null condition $k^{2}=0$ is relaxed, since $d\vphi(\tau)/d\tau  = s[\vphi(\tau)]\, k\cdot p/m$ the relation between phase and proper time becomes implicit and an integration is required to obtain the explicit relationship.
\newline

We have presented some solutions to the Klein Gordon equation for non-null ($k^{2}<0$), univariate transverse fields. The WKB solution ceases to be exact in this case, making the exact solution non-Volkov in character. Approximations based on WKB seem to work well, but can miss some of the non-perturbative structure e.g. the band-gap structure in parameter space for an electron in a periodic background field.
\newline

Non-null fields were mentioned to occur at the magnetic node of a standing wave, for which $k^{2}>0$. This is particularly relevant to simulations of electromagnetic cascades, which often invoke the (constant) plane wave model in a homogeneous, time-dependent electric field \cite{kirk08,ruhl10,elkina11,king13a}.
\newline

If a perturbative solution is invoked, in which the $k^{2}$ term is neglected, or included to first order by a ``reduction-of-order approach'' \cite{king16b}, for high-energy, over-the-barrier problems, the approximation appears promising (this depends on the field set-up: for example, see the problem at the magnetic node of a standing wave detailed in \cite{king16c}). The advantage with such an approximation is that it is independent of the form of the background. However, when barrier reflection or tunneling becomes relevant, important dynamical details are missed by all such first-order approaches.

\section{Acknowledgments}
B. K. acknowledges fruitful and productive work with A. Ilderton, T. Heinzl and H. Hu. B. K. is thankful for the support from the Royal Society International Exchanges Scheme, CARDC (Chinese Aerodynamics Research and Development Center) and of SWUST (Southwest University of Science and Technology), Sichuan, China.

\section*{References}
\bibliographystyle{iopart-num}
\bibliography{current}

\end{document}